\documentclass[preprint]{aastex}

\usepackage{psfig}

\def\deg{$^{\circ}\,$}
\def\solm{M$_{\odot}\,$}
\def\kms{km s$^{-1}$}
\def\ergs{erg s$^{-1}$}
\def\um{$\mu$m$\,$}

\begin{document}

\title{Feeding the monster... the nucleus of NGC~1097 
at sub-arcsec scale in the IR with the VLT}
\author{M. Almudena Prieto}
\affil{Max-Planck-Institute fuer Astronomie,  D-69117 Heidelberg, Germany}
\email{prieto@mpia.de}
\author{Witold Maciejewski}
\affil{Astrophysics, University of Oxford, Keble Rd., Oxford, OX1 3RH, UK}
\email{witold@astro.ox.ac.uk}
\and
\author{Juha Reunanen}
\affil{European Southern Observatory, D-85748 Garching, Germany}
\email{jreunane@eso.org}

\begin{abstract}
Near-infrared images of the prototype LINER / Seyfert 1 galaxy
NGC~1097 observed with the Very Large Telescope (VLT) using adaptive
optics disclose with unprecedented detail a complex central network of
filamentary structure spiralling down to the center of the galaxy. The
structure, consisting of several spiral arms, some almost completing a
revolution about the center, is most prominent within the radius of
about 300~pc. Gas and dust may be channelled to the center of NGC 1097
along this central spiral. Some filaments can be traced further out,
where they seem to connect with the nuclear star-forming ring at 0.7
kpc radius. Straight principal shocks running along the primary
large-scale bar of NGC~1097, seen in the optical images as prominent
dust lanes, curve into this ring, but radio polarization vectors cross
the nuclear ring under a rather large angle.  Here we attempt to
explain this morphology in terms of three-dimensional gas flow in a
barred galaxy. In our scenario, parts of the principal shock, which
propagate in the off-plane gas, can cross the nuclear star-forming
ring, and excite waves inward from it. If the dispersion relation of
the excited waves allows for their propagation, they will naturally
take the shape of the observed central spiral.  The nuclear region of
NGC~1097 remains unresolved at sub-arcsec scales in the near-IR, with
an upper size limit of $<$10 pc FWHM. Thus, any putative central dusty
torus or gaseosus disk envisaged by the AGN unified schemes has to be
smaller than 10 pc in diameter at near-IR wavelengths. The extinction
in the region between the nuclear star-forming ring and the nucleus
increases very moderately, reaching $ A_v \sim 1$ at the immediate
surrounding of the nucleus.  Thus, if the nuclear filaments are
tracing cold dust, they contribute to a very low extinction in the
line of sight and are likely to be distributed in a rather thin disk.

\end{abstract}

\keywords{galaxies: nuclei -- galaxies: Seyfert NGC~1097 -- infrared:galaxies}


\section{Introduction}

NGC 1097 is a SBb galaxy with an Active Galactic Nucleus (AGN), located at a
distance of 14.5 Mpc (Tully, 1988; 1~arcsec~$\sim$~70~pc), which places it
among the nearest AGN in the Southern Hemisphere. It has been originally
classified as a LINER and further reclassified as a Seyfert type 1 (Sy 1)
nucleus after the detection of broad, double peaked $H\alpha$ emission by
Storchi-Bergmann et al. (1993). However, it is very moderate AGN with
$L_{2-10 \rm{keV}}\sim 10^{40}$ \ergs (Terashima et al. 2002). Dust
extinction toward the nucleus, as inferred from the broad Balmer line ratio,
$H\alpha / H\beta = 3-4.2 $ (Storchi-Bergmann et al. 1995), is quite moderate,
$A_v \sim 1$. The ratio of Balmer 
narrow components indicates similar extinction (Storchi-Bergmann et al. 1996).

NGC 1097 has a very strong bar at $PA\sim$141\deg and a prominent nuclear 
star-forming ring of 0.7 kpc radius inside it. Interior to the  ring, a 
secondary nuclear
bar crosses the nucleus at PA=28\deg (Quillen et al. 1995), almost orthogonally
to the primary bar. The nucleus and the star-forming ring are prominent in CO 
and HCN molecular emission (Kohno et al. 2003).  The CO map shows three main 
peaks: at the nucleus and at two diametrically opposite locations in
the 
nuclear star-forming ring. The latter correspond to the connecting 
points between
principal dust lanes in the outer bar and the star-forming ring. They also 
roughly coincide with the ends of the inner nuclear bar, but they slightly 
trail the bar (have larger PA).  The resolution of Kohno's et al maps, $\sim
7.7''\times 4''$ in CO(1-0) and $10''\times4.5''$ in HCN(1-0), is insufficient
to confirm the presence of cold material between the nucleus and the ring. 
The map and radial profile distribution presented by the authors hint to 
the presence of some weaker CO emission in this region. First evidence for 
the presence of cold material there comes from the Hubble Space Telescope (HST)
optical images analyzed by Barth et al. (1995), who reported on the presence 
of dusty filaments in the surroundings of NGC 1097's nucleus. 

Regarding the presence of warmer gas in the nuclear region of NGC 1097, 
infrared (IR) spectroscopy reveals emission from both $H_2$ and HII at the 
nucleus, the star-forming ring and the region in between. Long-slit IR 
spectroscopy in several directions across the nucleus (Reunanen, Kotilainen \& 
Prieto 2002; Kotilainen et al. 2000) shows strong $H_2$ 2.12 \um line emission 
from the 
nucleus and the star-forming ring, and weak extended emission from the region 
in between. Optical long-slit spectroscopy (Storchi-Bergmann et al. 1996) 
shows strong ionized gas emission from both the nucleus (Sy 1 spectrum with 
broad $H\alpha$ emission) and the ring (low excitation HII region spectrum), 
and faint line emission from the region in between (a LINER type spectrum).  
Br$\gamma$ emission is detected in the ring but not at the nucleus (Reunanen, 
Kotilainen \& Prieto, 2002; Kotilainen et al. 2000). 
Also the $H\beta$ line is 
very weak in the nuclear spectra due to the absorption component by 
 the underlying 
stellar contribution.

In this paper we present high spatial resolution images of NGC1097, obtained 
with the Very Large Telescope (VLT) and the Adaptive-Optics-assisted NACO 
camera/spectrograph in 
the 1-2.4 \um range. The physical scales traced by these images are of the 
order of 10 pc, and thus they probe with unprecedented detail the presence and
extent of obscuring material in the very proximity of the nucleus.
After describing the observations in Section 2, and results in Section 3, in 
Section 4 we attempt to explain the dynamics of the  
central filamentary structure that is observed.

\section{Observations}

Images of the NGC~1097 nucleus in $J-$ $H-$ and $Ks-$ bands were collected 
with the UT4 unit of the VLT during August 2002.  The images were taken with 
the NACO camera that provides a field of view of 27$\times$27 arcsec and a 
scale of 0.027 arcsec per pixel.  The adaptive optics correction was done
with the optical wave-front sensor closing the loop on NGC~1097 nucleus.
The achieved spatial resolution has FWHM of about $0.18''\times 0.15''$ 
in $Ks-$band, $0.20''\times 0.18''$ in $H-$band, and $0.20''\times0.19''$
in $J-$band. These values were measured on various isolated HII regions in 
the star-forming ring, which we assumed to be point-like. These HII regions 
are resolved in the HST WFPC1 image at 5500~\AA$\,$ (Barth et al. 1995), having 
sizes about 4 pc (0.06 arcsec). In the NACO images,  
we  get almost constant angular 
size for several of the HII regions, 
as well as for the nucleus, and thus we believe this to
be the real resolution of the data. It is different from  that derived from 
a stellar PSF, but
estimating the spatial resolution on the basis of separate images of 
stars is not viable because atmospheric correction achieved by the Adaptive
Optics (AO) system 
is different depending on the type of source used for monitoring the
atmosphere. NGC~1097 images were corrected using the nucleus of the galaxy,
which means that the contamination by the galaxy light degrades the contrast 
of the nuclear light. This leads to a systematically worse AO correction when
compared with that achieved from a separate image of an isolated bright star.
The spatial resolution achieved in the PSF star images
taken for this program turned out in general to be at least 30\% better than
that of the science images.

NACO $J-$ and $Ks-$band images, and the HST WFPC1 optical image of NGC 1097 are
shown in Fig. 1. The NACO $H-$band image is similar, and therefore it is not 
shown. Although the resolution of the NACO images is about 4 times worse than
that of the optical HST image, this shortcoming is largely compensated
by the amount of new information revealed in the IR images.  At all
wavelengths the AGN  and the star-forming ring are detected, but
the IR images show in addition the presence of a central nuclear bar
and newly discovered HII regions in the ring, more than 300 compared
with 88 measured by Barth et al. (1995) in the HST image.  The
comparison of the optical and IR images gives evidence for the presence of dust
all over the star-forming ring interlaced with the HII regions.  The
ring morphology is more like two spiral arms, with each of the arms connecting 
to an end of the nuclear bar. Note however that a different morphology is 
disclosed by the $J-Ks$ color image presented in Section 3.3.

\section{Results}

\subsection{The nucleus}

At the resolution of these IR observations, the nucleus of
NGC~1097 remains a point-like source, with an upper limit to the size of  
$0.19'' \times 0.14''$ FWHM in the $Ks-$band.  
This is about the FWHM measured in
several of the best isolated HII regions in the star-forming 
ring from the same $Ks$-band 
image (see Section 2), and thus we consider the nucleus unresolved.  At the
distance of NGC 1097, that angular resolution implies an upper limit
for the nucleus size of less than 10 pc diameter.  
 
A light-profile decomposition of the central 3 arcsec ($\sim$210 pc) 
radius of 
NGC~1097 is shown in Fig. 2. It is fitted with a simple 
two-component model consisting of an unresolved source -- the nucleus, represented 
by a PSF,
and an underlying galaxy component or pseudo-bulge, 
represented by a generalized Sersic model.
Surface brightness in the Sersic model follows a formula
$I(r) = I_o \exp [-(r/r_o)^{1/n}]$, 
where $I_o$ is the central surface brightness, $r_o$ is a scaling radius, 
and $n$ is the Sersic exponent (Sersic 1968). The PSF was estimated as
follows. The useful dynamic
range of the profile from the HII regions in the star-forming ring 
 is only $\sim5^m$; the
dynamic range reached in the PSF star images observed separately from
the science frames is however $10^m$.  The core of PSF star images is
narrower than that measured for the HII regions. Therefore, a PSF
composite profile was produced which includes the profile of an HII
region up to a radius of 0.45 arcsec in $Ks-$band, 0.3 arcsec in $J-$ and
$H-$ bands, complemented with that of the stellar PSF from those radii on.
The Sersic exponent $n$ was allowed to vary between 1 and 4, but the best 
fit in any of the bands was obtained for $n=2$. The results of the best fit
are shown in Fig. 2, and the parameters derived from this
fit are given in Table 1. The residuals from the fit are less than 5\%
at all radii. However, within 0.3-0.6 arcsec radius, the residuals
show structure which is most probably due to a still unsatisfactory
PSF, particularly in $J-$ and $H-$ bands.

The contribution of the point-like source, the AGN, to the total light at different
radii is given in Table 2.  As expected, the AGN
contribution increases with wavelength.  Within a 0.2 arcsec aperture
diameter, which is $\sim$ 1.5 times the achieved spatial resolution,
its contribution  in the $Ks$-band is dominant, 90\%, but
it is 20\% less important in $J-$band.  The colors of the nucleus are
rather red: $J-H=1.3$, $H-Ks=1$, indicative of dust temperature of 1000
K. However, this temperature should be taken as an upper limit
considering that the measurements still correspond to relatively large
physical scales --- the resolution of these
observations is about 10 pc. Warm and hot gas should also be present
at these distances from the center and therefore 
  free-free emission from these gas phases might  also be
contributing at those scales to the near-IR, in particular if shock
excitation is present. This indeed seems to be the case as indicated
by the large line-widths, $100-300$ \kms, measured in the optical
spectra from  regions in between the nucleus and the star-forming
ring (Storchi-Bergmann et al. 1996).  Shock
velocities of that range can heat the gas to $T\sim 10^6 K$, and
bremsstrahlung emission from this gas will substantially contribute in
the IR (see e.g. Contini et al. 2004).

\subsection{The pseudo-bulge}

Table 2 gives the colors of the region between the nucleus and the 
star-forming ring integrated directly on the images over increasing 
concentric radii. This region, 0.7 kpc in radius, is referred to as 
the pseudo-bulge of NGC 1097 in Section 3.1. The colors
are subtracted  from the point source contribution estimated from the
radial profile fit (Section 3.1). Equivalent colors derived from the Sersic model 
used in the fit are given in  Table 2 for comparison. There is reasonable 
agreement within a few percent
between the two estimates except for radii below 0.4 arcsec where  the
nucleus and an additional source of emission 
 become important. This is discussed below.

The radial variation of the colors in the central 10 arcsec  together
with a color-color diagram from the same region are plotted in 
Fig. 3. These colors 
are measured on the images from
azimuthally averaged concentric rings.  By comparing the upper and
lower panels in Fig. 3, one sees that the reddest
colors, i.e.  $0.8 < J-H< 0.95 $ and $0.3< H-Ks < 0.5$, are measured
at radii below 1 arcsec.  The particular trend
followed by these extreme colors in the $J-H$ vs $H-Ks$ diagram is caused
by contamination by the nucleus light particularly in $Ks-$band (see Table
2). It may also be introduced by the underlying central  spiral
structure revealed in the color images (see Section 3.3), which
becomes more dense with decreasing radius.  This contamination seems
to fade at about 1 arcsec ($\sim$70 pc) radius as indicated by the
clear color turnover at $J-H \sim 0.8$ and $H-Ks \sim 0.3$ in the color-color
diagram. From that radius outward pseudo-bulge dominates the color,
which becomes  monotonically bluer, reaching a minimum $J-H \sim 0.7$,
$J-Ks \sim 0.24$ at $\sim 5$ arcsec (350 pc) radius from the
center. Further out, the colors become redder again due to the
incipient contribution of NGC 1097 nuclear star-forming ring.

With the aim of determining the extinction towards the nucleus, 
as a first step we took 
those minima, $J-H \sim 0.7$,
$J-Ks \sim 0.24$, as the intrinsic colors of  NGC 1097 pseudo-bulge. In doing
so, we note that the progressive reddening with decreasing
radius is difficult to explain by dust extinction alone,
unless very steep IR extinction curves, with $\alpha \sim$ 3.5
($A_\lambda \propto \lambda^{-\alpha}$), are considered. The problem is
better illustrated in Fig. 4 where a detailed view of the color
variations in a selected area of the bulge -- e.g. the South-West
quadrant -- is shown.  $J$-$H$ and $H$-$Ks$ are measured in
consecutive boxes of size 0.27$\times$0.27 arcsec, covering the region between
0.5 and 5.5 arcsec ($\sim 40 - 400$ pc) from the center.
Colors measured in the other quadrants show similar behavior.  Taking
again $J-H \sim 0.7$ and $J-Ks \sim 0.24$ as the intrinsic colors of the
pseudo-bulge, the progressive reddening at radii larger than $\sim 3$ arcsec 
is well explained by dust extinction alone (two extinction laws
with $\alpha=1.5$ and 1.8 are shown in the figure).  However further in,
$J-H$ becomes much redder than predicted.

We then compared the observed color variation with the one predicted  by 
the Sersic model used to fit the pseudo-bulge (Section 3.1). These colors are 
plotted as a red line in  Fig. 4. It can be seen that the observed 
colors follow Sersic predictions down to a radius of $\sim 2.5$ arcsec. 
Further in, they depart from Sersic trend, but the departure is now easily
reconciled with a more pure extinction law. Based on this result, we estimate
the extinction toward the nucleus by taking as a face value for the
intrinsic colors of the pseudo-bulge those measured at $\sim$2.5 arcsec
(180 pc) radius from the center: i.e. $J-H\sim0.78$, $ J-Ks\sim$ 0.25
(Fig. 4).  The reddest values at about 0.6'' from the nucleus 
are $J-H \sim 0.86$, $J-Ks \sim 0.32$. These compared with our fiducial
value lead to a moderate extinction of $A_v\sim 1$ within the
central $\sim 60 - 40$ pc region. This extinction is in the range
measured from the nuclear Balmer decrement estimated from an aperture
 larger than 1 arcsec (Storchi-Bergmann et al. 1996).

\subsection{The central spiral network}

In  Fig. 5 (bottom-left panel),    a $J$-$Ks$ color map of the central 
2 kpc region of NGC~1097 is presented. It shows the morphology 
of the nuclear 
star-forming ring as a complete annulus, filled by a continuous distribution 
of HII regions and gas. It differs from two spiral arms morphology 
seen in the optical 
image (Fig.~1). In the $J$-$Ks$ color map most of the contribution 
from both the pseudo-bulge and the nuclear bar cancels out, disclosing a
filamentary structure that spirals around the center. This structure is 
similar to those unveiled in nearby galaxies by 'structure maps' constructed
from HST WFPC2 images (Pogge \& Martini 2002), or through visible-near-infrared
color maps from the HST (Martini et al. 2003). The resolution of 
our NACO images, when compared to that of HST/NICMOS images, 
is similar in $H-$band and 
a factor of 2 worse in $J-$band. However, the PSF of the VLT/NACO images 
does not show the boxy diffraction pattern often seen in NICMOS images of
strong point-like sources, and therefore nuclear structure  at distances 
as small as 10 pc from the center can be revealed. An additional  advantage of
using a near-IR color image in this study is that this spectral range is 
less prone to the contamination by line emission. $V-H$ images often used  
to trace dust structures in the surrounding of an AGN may largely be 
polluted by strong optical emission lines  in the $V-$band. 

Most of the central filamentary structure around the nucleus of NGC~1097 
cannot be distinguished in the 
direct HST optical and NACO IR images, except for the outer parts of the 
longer filaments which are seen as obscured fingers in the HST WFPC1/F5500W
and the NACO 
$J-$band images (Fig.~1). These filaments are seen in extinction against 
background light from the pseudo-bulge, and therefore they are likely to be 
tracers of cold material and dust. The low extinction, $A_v \sim 1$,
indicated by the IR colors of the pseudo-bulge (see Section 3.2) suggests that the
filaments are not filling much of the volume within this region, and are likely
distributed in a rather thin disk.

To investigate the central filamentary structure in more detail, a simple 
elliptical model was fitted and subtracted from the $J$-band image, with 
the fit restricted to the region inside the nuclear star-forming ring. 
The residual image, shown in  Fig. 5 (bottom-right panel),  
looks like the negative of the $J$-$Ks$ 
color map. The brighter filaments in the color map appear as dark
channels in the residual, yet, with sharper detail and increasing length in 
the latter. Our analysis of the VLT/NACO images of NGC~1097 shows that these 
filaments, within the achieved resolution of about 10 pc, end up at the very 
center of the galaxy. Several filaments can be traced in the residual image,
but most prominent are three spiral arms that wind around the center at radii 
below 200 pc. At larger radii, the pitch angles of these 
arms increase, and two arms seem to align with the nuclear secondary 
bar present in NGC~1097. The northern arm then splits into two at a
radius of about 300 pc. The third arm appears to be unrelated to the 
nuclear bar, and it splits into a number of spiral filaments at a similar 
radius. Some of the filaments can be followed outward up to the 
nuclear
star-forming ring. It can be seen in the HST/ACS image (Fig. 5, top panel) that
this ring also encircles the curving innermost parts of the dust lanes which 
run along the primary bar of NGC~1097. The nuclear filaments seem to connect 
with the large-scale dust lanes, but there is a clear discontinuity in the 
pitch angle at the meeting points.

\section{Dynamics of the central spiral network}

The VLT high-resolution infrared images presented in this paper unveil a
complex spiral network of dust filaments around the nucleus of NGC~1097. 
Here we examine how such
a network could have been formed, and whether it could funnel gas inward. 
The mass of the molecular gas makes only a fraction of the total
dynamical mass within the inner few hundreds parsecs of the galaxy. 
We estimate an upper limit to the central
molecular gas mass within the beam size of the CO observations by Kohno et
al.  (2003), $\sim 7.7''\times 4''$ or $\sim 550 \times 270$ pc, to be
$\sim 5 \times 10^{7}$ \solm. The dynamical mass within the same area,
inferred from the rotation curve (Storchi-Bergman et al. 1996), is at
least $\sim 5 \times 10^8$ \solm.

\subsection{Origin of the central spiral}

The spiral dusty filaments around the nucleus of NGC~1097 are unlikely to be
formed by a self-amplified density wave, analogous to the one that shapes
the classical large-scale spiral arms in galaxies. This is because stars,
which make up most of the mass in this region, mostly settle in the pseudo-bulge 
and in the nuclear bar that we observe in the infrared. This spiral pattern 
is also unlikely to be of
acoustic origin (Elmegreen 1994), because it consists of a few sharp
spiral filaments, while acoustic spirals, generated by the amplification
of sound waves in the gaseous nuclear disk, are expected to have
flocculent, multi-arm morphology, and to uniformly fill the nuclear disk.
The spiral network around the nucleus of NGC~1097 closely resembles the
nuclear spirals seen in hydrodynamical models (Englmaier \& Shlosman 2000,
Maciejewski et al. 2002, Maciejewski 2004b). Straight principal shocks in 
a large-scale bar may end in a nuclear ring, which itself is a tightly wound
spiral, or they can give rise to more open nuclear spirals, which are
an extension of these shocks inward to smaller radii. Hereafter we call
the nuclear spirals that extend to the galactic center the {\it central 
spirals}. 

There is a strong large-scale bar in NGC~1097, which could
trigger a nuclear spiral,
 but there is also a prominent nuclear star-forming ring 
there, abundant in gas. Central spirals have not been found in models
inside gaseous nuclear 
rings, because shocks triggered by the large-scale bar get 
damped in such rings, so they cannot give rise to central spirals
(Maciejewski 2004b). The coexistence of the nuclear star-forming ring 
and the dusty central spiral in NGC~1097 seems to contradict these models, 
unless principal shocks in the bar have some effect on the gas encircled 
by the nuclear ring.

Numerical models of gas flow in bars have so far 
only been built in two dimensions, in order to reflect the dynamics of gas
in the disk of a galaxy. On the other hand, shocks generated in gas by a
large-scale bar are intrinsically three-dimensional, and they propagate
also in the off-plane gas. There have been studies of the vertical structure
of galactic spiral arms (Martos \& Cox 1998, Gomez \& Cox 2002), but not
of the shocks in bars. Here we make a qualitative attempt to describe
gas flow in bars in three dimensions. We base our description on two
assumptions: 1) in addition to the gas in the plane of the disk there is
off-plane gas of lower density, 2) larger scale-height of the off-plane gas
is supported by its energy stored in velocity dispersion of the clouds.
Within this framework we can guide our understanding of gas flow in three 
dimensions using various two-dimensional models for various heights above 
the galactic plane. However, full quantitative description requires 
high-resolution three-dimensional hydrodynamical models, which are beyond
the scope of this paper.

The dispersion relation of a wave generated in the gaseous disk implies that
the pitch angle of the wave is proportional to the velocity dispersion in gas
(Englmaier \& Shlosman 2000, Maciejewski 2004a). Thus in gas with higher
velocity dispersion (dynamically warmer), waves generated by the principal 
shocks in the bar should have higher pitch angle than in a dynamically colder 
gas. When the pitch angle is high enough, waves can spiral freely into the 
galactic center (Maciejewski et al. 2002). This may be a scenario 
for the extraplanar gas. On the other hand, shocks propagating in the disk,
where gas is dynamically colder, generate waves with smaller pitch angle. These
waves are damped in dense post-shock gas, and a nuclear gaseous ring forms with
excellent conditions for star formation (Maciejewski 2004b). The morphology of 
waves, which are expected to form in the disk as well as in the extra-planar 
gas, together with gas morphology in the plane of galaxy are shown in the top 
panel of Fig. 6. These morphologies should be compared with those in the 
bottom-right panel of Fig.5. Although the structure in the model is generic,
and not tuned to reflect particular features in NGC~1097, there are 
similarities between the model and the data. The large-scale dust lanes enter
the nuclear star-forming ring along the paths marked by the red lines that 
indicate shocks in the cold in-plane gas in the model, whereas the observed
central spiral takes a shape similar to that of the green line marking
shocks in the extraplanar gas. The distribution of the emission from the
star-forming ring resembles in-plane gas density distribution in the model.

The nuclear star-forming ring in NGC~1097 consists mostly of
molecular gas, and therefore it is well confined to the galactic plane. It
damps waves generated by the principal shocks in the bar that are propagating 
in the galactic plane.  However, in the scenario proposed above, shocks
propagating in the off-plane gas may pass the ring almost unaffected. 
Inward from the ring our simple plane-parallel scenario breaks down, and
full three-dimensional approach is needed to determine whether off-plane
shocks couple with the dynamically cold gas in the disk and whether they
can generate spiral density 
waves there. It is possible that the gaseous disk inside the nuclear 
star-forming ring
is no longer dynamically cold, as indicated by the large pitch angle of the
central spiral, which should facilitate the coupling. If the dispersion 
relation of the excited wave allows for its propagation inward from the
nuclear ring, this wave will naturally take a spiral shape. Waves
inward from the ring can propagate freely toward the galactic center,
because they are not hindered by the dense post-shock gas in the galactic 
plane, which has settled in the ring outside them. We hypothesize that these 
waves form the observed spiral network inside the nuclear star-forming
ring. In this picture the central spiral in NGC~1097 would be a wave in 
gas, and a continuation of the principal shocks present in the large-scale 
bar.

We note that the innermost parts of the central spiral have a three-arm
symmetry, not an obvious consequence of triggering by bisymmetric
shocks in a bar. Three-arm spirals can form as a response to lopsided ($m=1$)
or $m=3$ terms in the forcing. One source of these odd terms in NGC~1097 may be
the difference in strength of the two principal shocks in the main bar.
It has been observed 
that the amount of post-shock gas can differ between the two shocks by a 
factor of two (Schinnerer et al. 2002). This can effectively introduce
odd terms in forcing, since in our scenario it is the shock, and not the bar, 
that triggers the central spiral. A dwarf elliptical companion to NGC~1097
may also introduce $m=1$ mode in the shocks. Regardless of its explanation, 
the three-arm symmetry of the central spiral indicates that there is a 
departure from bisymmetry in the forcing that generates this pattern.

\subsection{Interpretation of the polarization maps of NGC~1097}

Our scenario for the origin of the central spiral in NGC~1097, presented
above, is supported by the morphology of the regular magnetic field seen in 
the radio polarization maps of this galaxy (Beck et al. 1999). In the lower
panel of Fig. 6, we present the magnetic field vectors observed by Beck et al. 
(1999, 2005), which are aligned with the shearing flow, plotted on top of our 
processed $J$-band image that shows the dusty central spiral. At radii where 
the principal shocks originating in the main bar approach the nuclear 
star-forming ring, the magnetic vectors progressively swing from an alignment 
with these shocks to a clear spiral pattern that crosses the nuclear star-forming ring under
a rather large angle ($\sim$50\deg). Directly inward from the ring this
pattern tends to follow central dusty spiral structure that we
observe.

At the position of the nuclear star-forming
 ring, the spiral morphology of the magnetic 
field is very different from the ring-like morphology of gas and of 
star-forming regions. Here we propose that this is because star formation is 
confined within the 
disk of the galaxy, while radio polarization maps show the regular magnetic 
field averaged along the path length that includes the off-plane gas. The 
scale-height of the off-plane gas, from which polarized emission is observed
in galaxies, is about 1 kpc, as
indicated by maps of the magnetic field in edge-on spirals (Beck 2000).
Although the total radio signal is largest from the dense in-plane gas,
star-forming activity creates field turbulence there, which lowers the 
degree of polarization in the nuclear star-forming ring in the plane of 
NGC~1097. If field turbulence in regions far from the galactic plane is much 
lower, then most of the polarized emission may be coming from the off-plane 
gas, even when most of the total radio emission is coming from the disk.
Then vectors in polarization maps will trace kinematics of the 
off-plane gas. Polarization vectors observed in NGC~1097 (bottom
panel of Fig.6) indicate strong ordered magnetic field on top of the 
star-forming nuclear ring, as well as a field aligned with the large-scale
dust lanes and with the central spiral. Directions of these vectors
are consistent with our scenario, in which parts of the principal shock that
propagate in the off-plane gas give rise to a spiral pattern, 
which crosses the nuclear ring. Note that in this scenario the magnetic 
field is just a tracer of gas motions, and it does not play a dynamical role.
The role of magnetic fields in extraplanar gas is most likely important, and
full magnetohydrodynamical approach may be needed in order to understand
the propagation of waves past the nuclear ring.

If the polarized emission comes mostly from the off-plane gas, then another 
paradox in NGC~1097 noticed by Beck et al. (1999) can be explained in a 
similar way. Beck et al. observe a depolarization zone, which they ascribe to 
a shock, {\it upstream} from the straight principal dust lanes. Hydrodynamical 
models (e.g. Englmaier \& Gerhard 1997, Maciejewski et al. 2002) indicate that 
the distance of the principal shock from the major axis of the stellar bar 
decreases with 
the velocity dispersion in gas. This is also seen in the top panel of Fig. 6. 
If the depolarization zones mark shocks in the off-plane warmer gas, then they 
are expected to be located closer to the major axis of the bar than the dust 
lanes, which mark shocks in the dynamically cold gas in the plane of the disk.

\subsection{Central spiral in NGC~1097 as a shock in gas. The role of the
inner bar}
Whether the central spiral in NGC~1097 is a shock or a weak wave has direct 
implications for the expected inflow of gas to the center. A weak wave 
triggers no inflow, while shocks are capable of supplying enough gas to the 
galactic center to maintain the AGN activity (Maciejewski 2004b). Nuclear 
spirals in barred galaxies can be shocks throughout their extent from the 
principal shock in the bar to the galactic center (Maciejewski et al. 2002).
Nuclear spiral shocks in gas tend to have pitch angle larger than what the 
linear theory of gaseous density waves predicts (Maciejewski 2004b).  
The observed large pitch angle of the central
spiral in NGC~1097 is beyond the linear regime, which is suggestive of 
a spiral shock. The optical emission from the region between the
nucleus and the ring reveals a typical LINER spectrum with strong low
excitation lines and large line widths (FWHM of [NII] 6548 \AA\ between
100 and 300 \kms, Storchi-Bergmann et al. 1996), which are signatures of
shock-excited gas (Contini \& Viegas 2001, Viegas \& Contini, 1989). This,
together with the observed lack of star formation in the central spiral,
favour the idea that this spiral is a shock with strong shear.

In Section 3.3 we noted that two of the three most prominent arms of the
central spiral in NGC~1097 unwind at radii above 200 pc to almost straight
filaments aligned with the inner bar. Thus one may think of an alternative
explanation, which ascribes this spiral to the principal shock in the
inner bar. However, principal shocks triggered by bars approach the major
axis of the bar toward the bar's ends, as seen both in nature and in numerical 
models, and accounted for by studies of orbital structure. On the contrary, 
in NGC~1097 the most pronounced north filament running along the inner
bar departs from the bar's major axis as the radius increases (see NACO 
$J$-band image in Fig. 1). This difference already indicates that the gas 
dynamics in the inner bar of NGC~1097 is different from the well studied 
cases of large-scale bars. In particular the central spiral in this galaxy, 
with its three-arm symmetry, is most likely not generated by the inner bar. 
Nevertheless, appropriate modelling of this dynamics is necessary before 
firm conclusions are to be drawn about the role of the inner bar. The only 
models of a spiral density wave propagating through a nuclear bar built so
far (Maciejewski 2004b) indicate that this bar can reshape the spiral shock,
generated there by the outer bar. It does so by straightening parts of the 
spiral arms and aligning them with itself, which gives an impression of 
straight shocks, and resembles the morphology of the two spiral arms observed 
in the center of NGC~1097. However, the nuclear bar in those models is rather 
strong, while the inner bar observed in NGC~1097 is weak.

Finally, the curling of the spiral pattern in the innermost 100 pc (its
pitch angle decreasing inward) may point at the presence of a
supermassive black hole in the center of NGC~1097. Such a black hole in
the center of a galaxy causes the nuclear spiral to wind up as it
approaches the center, while in its absence the spiral would be unwinding
with decreasing radius (Maciejewski 2004a).

\section{Conclusions}

The nucleus of NGC~1097 remains a point-like source at subarcsec scales in 
the near-IR. An upper limit to its size has FWHM $<10$ pc. Accordingly, any
putative central dusty torus or  gaseous disk required by AGN unified
schemes  has to be smaller than 10 pc in diameter at 
near-IR wavelengths.

High spatial resolution achieved in the NACO/VLT observations 
of NGC~1097 presented here was sufficient to reveal in sharp detail an 
intricate network of filamentary structure in the central 10'' (700 pc) 
of the galaxy. These filaments are seen directly in $J-Ks$ color maps
(Fig. 5, bottom left). They also appear in extinction 
against the IR background light from the galaxy, 
and therefore they are likely
to be tracers of cold material. On the basis of the IR colors, and assuming 
foreground extinction, a very moderate extinction toward the nuclear region
($A_v\sim < 1$) is determined. This is consistent with the extinction values 
derived from the Balmer decrement. Thus, if the filaments are tracing cold
dust, they contribute to a very low extinction in the line of sight, and 
therefore they are likely to be distributed in a rather thin disk. The 
filaments are spiralling around the center from the nuclear star-forming 
ring to the nucleus, 
which is suggestive of them being channels by which cold dust and gas 
are flowing to the center. The CO and HCN maps hint to the presence of cold
material in this region, though their resolution is insufficient to confirm
the exact location of this material. The molecular $H_2$ 2.12 \um emission 
is also present between the nucleus and the star-forming ring. We have
recently collected AO-corrected  $K-$band spectra of this region
which should resolve spatially the dynamics of this $H_2$  warm gas.

The spiral network in NGC~1097 resembles nuclear spirals found in
hydrodynamical models, where they are density waves driven by the shocks 
generated in a large-scale bar. There is a strong large-scale bar in NGC~1097 
(Fig. 5, top panel), which can generate the shocks, but the nuclear 
star-forming ring may hamper propagation of the waves inward in the plane of 
the disk. In this paper we propose a scenario, in which, because of different 
dynamical conditions in the off-plane gas, waves propagating there wind into a 
spiral, and pass the nuclear ring. We cannot make a decisive statement on how 
these waves propagate inward from the nuclear ring -- this would require full
three-dimensional and probably magnetohydrodynamical models -- but if the 
dispersion relation of waves excited there allows for their propagation,
they should naturally form the observed central spiral. Radio 
polarization maps of NGC~1097 support this scenario, on the assumption that the
polarization vectors trace the motion of the off-plane gas. These 
vectors show a progressive swing from an alignment
with the principal shocks along the primary bar to a clear spiral
pattern at the ring and further in. Interior to the ring, they closely
follow some of the nuclear filaments (Fig. 6, bottom panel). Emission lines
in this region, which show all the  
characteristics of shock-excited gas, together with large pitch angle of the
central spiral, and the lack of star formation in the spiral arms, make a
suggestive evidence that this spiral is a shock with strong shear, along
which gas and dust can be channelled to the center of NGC~1097.

\section{Acknowledgments}
We are thankful to Rainer Beck for interesting discussions and for providing 
us with the radio polarization map of NGC~1097 displayed in Fig. 6, bottom. We acknowledge comments 
from John Maggorian on this manuscript. This work was partially
supported by the grant 1 P03D 007 26 from the Polish Committee for Scientific 
Research.

\clearpage
\begin{figure}
\includegraphics[width=.5\textwidth]{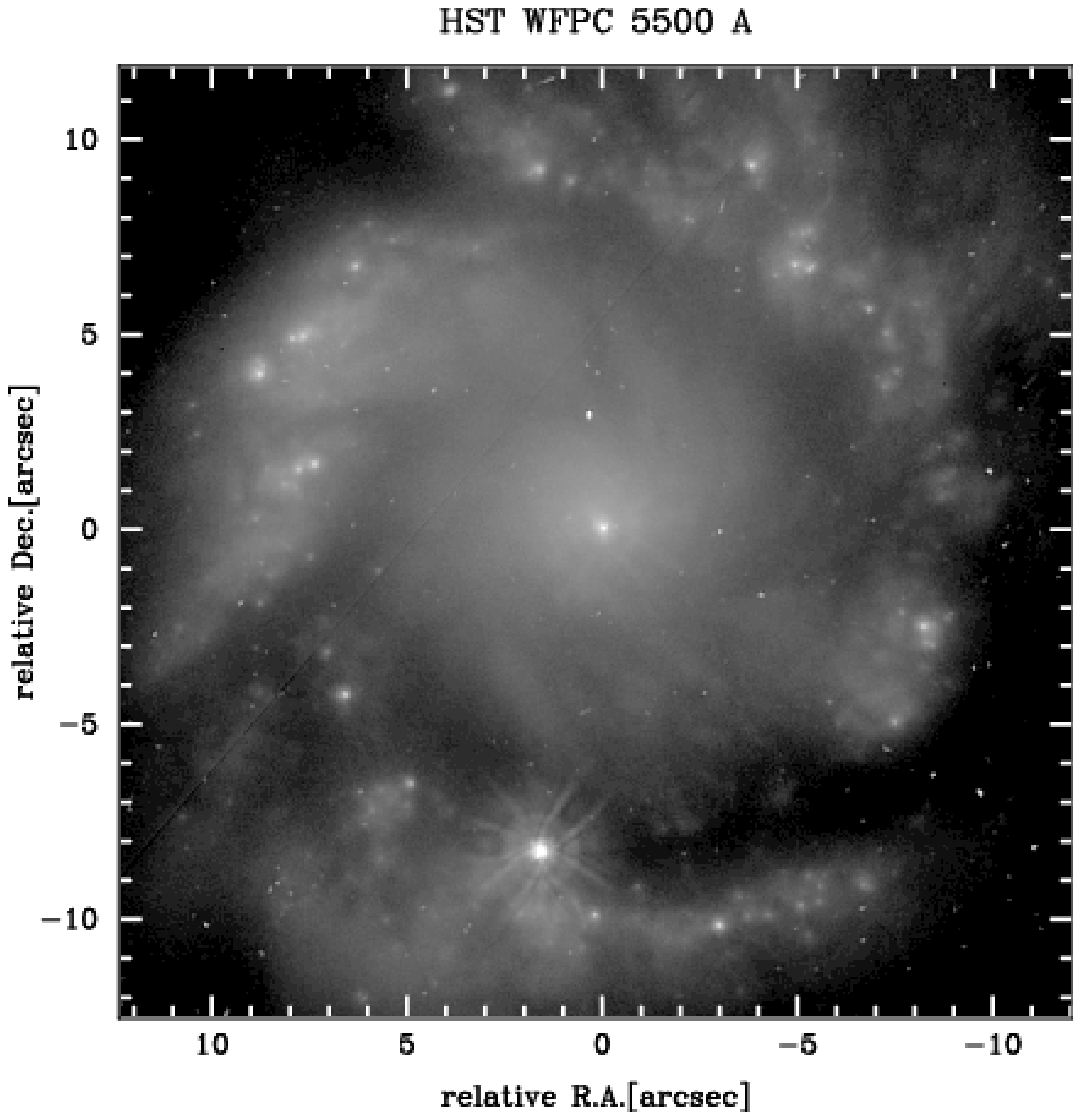}
\includegraphics[width=.5\textwidth]{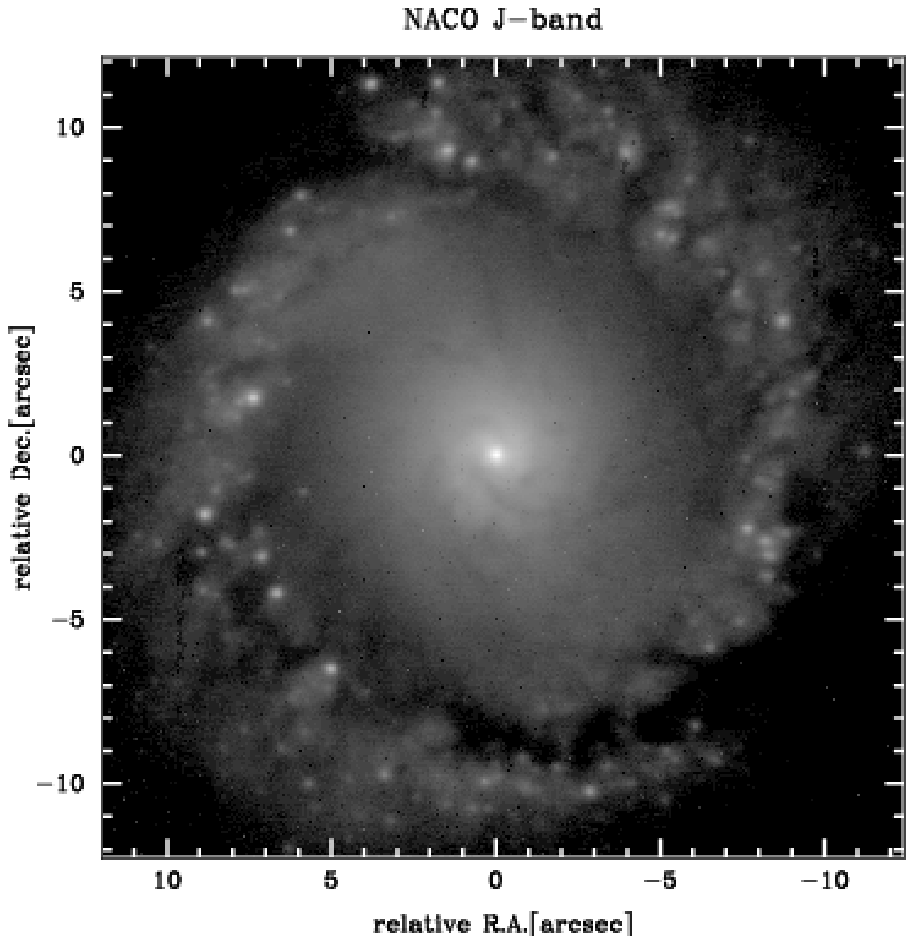}
\vspace{0.cm}\includegraphics[width=.5\textwidth]{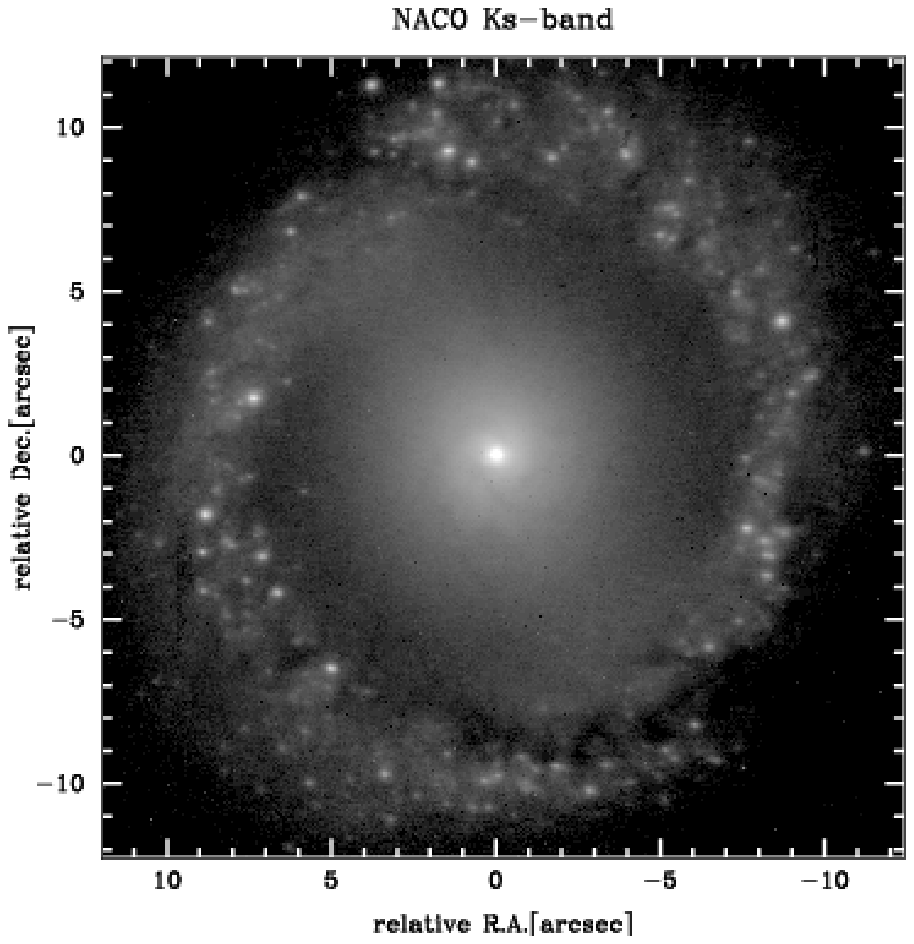}
\vspace{0.cm}\includegraphics[width=0.5\textwidth]{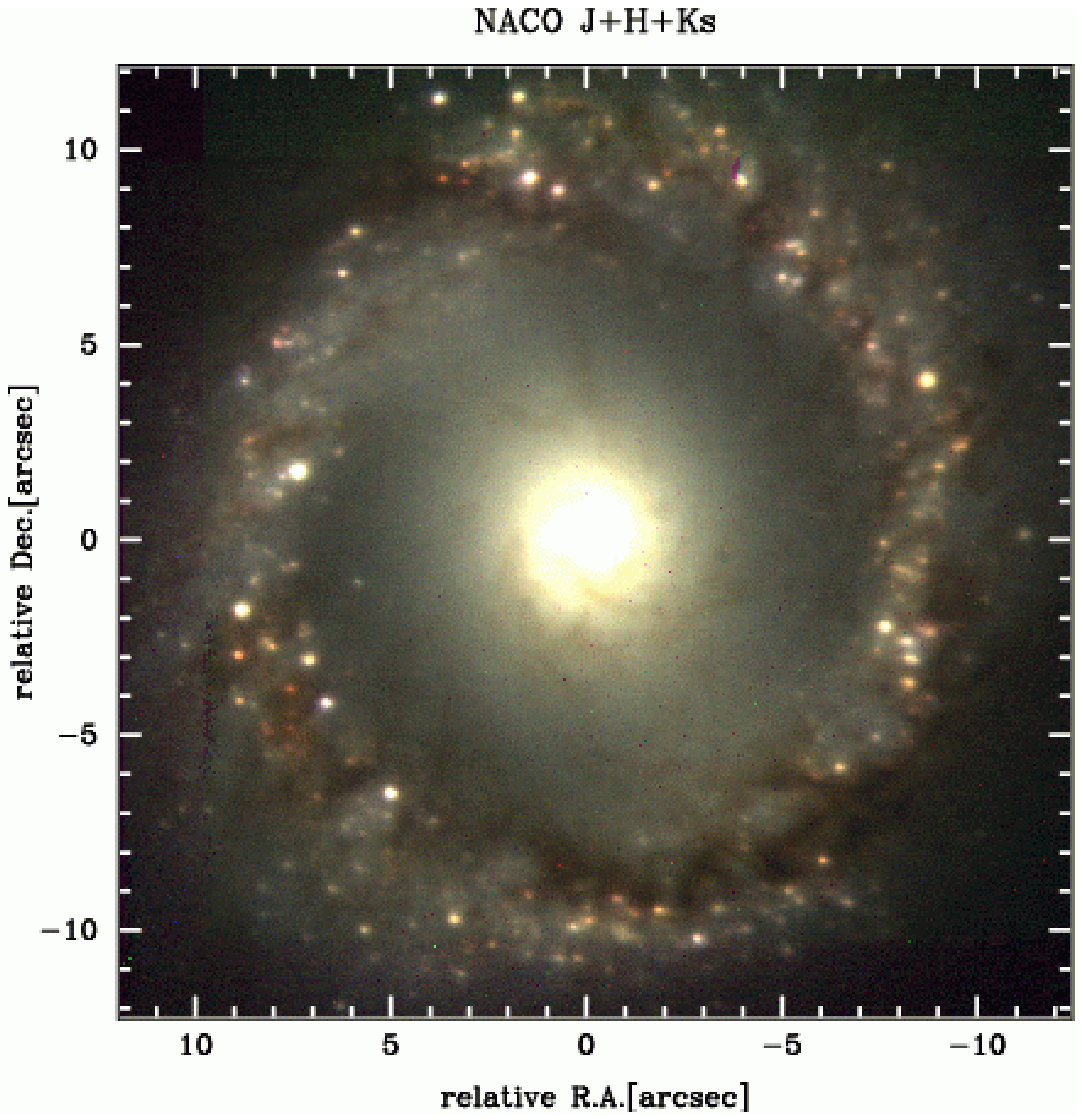}
\caption{Images of the central $\sim1.7 \times 1.7$ kpc region of NGC~1097 
taken with HST WFPC1 in the 5500~ \AA band (top-left), and with NACO in $J-$ 
(top-right) and $Ks-$ (bottom-left) bands. The grey-scale is logarithmic,
and cuts are chosen to emphasize the nuclear point-like source. The lowest
grey level  displayed in the three images is $\sim 2 \sigma$ over the background noise. 
The nucleus is  $\sim 40 \sigma$ in the HST and NACO $J-$band 
images, and  $\sim 150 \sigma$ in the $Ks-$band image. The average 
level in the region between the nucleus and the star-forming ring is $\sim 5 \sigma$ in HST 
and $J-$band images, $\sim 7 \sigma$ in the $Ks-$band image.
Bottom-right:  NACO true color image of the same field, constructed by 
stacking 
$J-$band (blue), $H-$band (green), and $Ks-$band (red) images. Color cuts 
are selected to saturate the nucleus but emphasize the star-forming ring. 
North is up, East is to the left in all panels.}
\end{figure}

\begin{figure}
\resizebox{0.7\hsize}{!}{\includegraphics{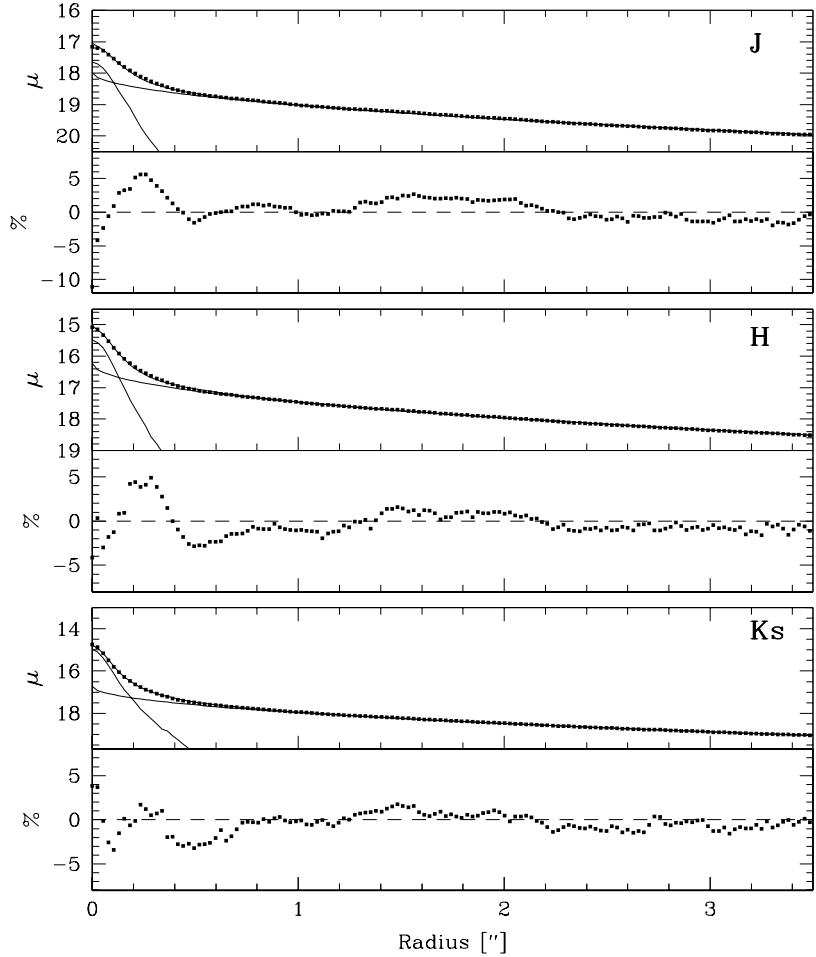}}  
\caption{Surface brightness profiles, in magnitude per square arcsec,
of the 
central 3 arcsec ($\sim$270 pc) region of NGC~1097, in the three NACO bands. 
The line running through the data points is the fit to the observed profile. 
It is based on two components: a point source and a Sersic model with $n=2$, 
both shown  in the plot. The residuals from the fit are  
 appended below each fit.}\label{radial}
\end{figure}
\clearpage
\begin{figure}
\resizebox{0.6\hsize}{!}{\includegraphics{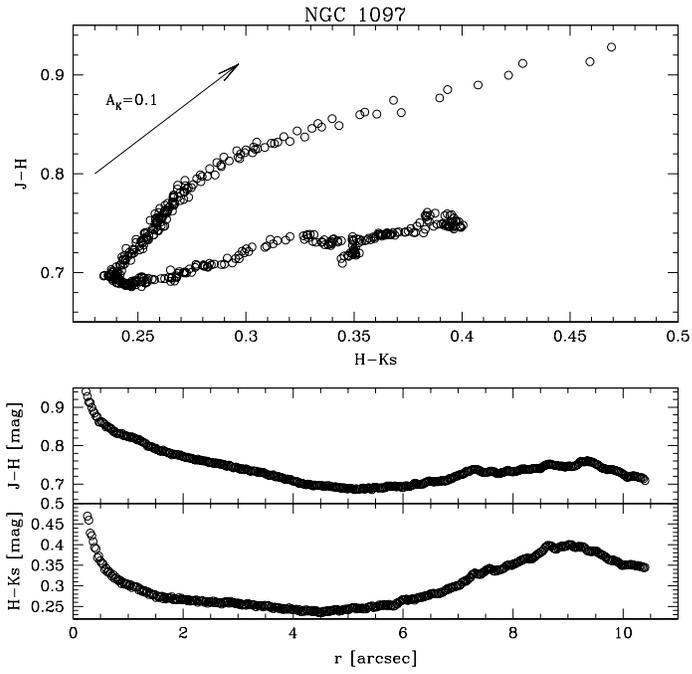}}
\caption{{\bf Upper panel:} $J-H$ vs $H-Ks$ diagram for the central 1 kpc 
region of NGC~1097. The reddening law direction is indicated by the arrow.
{\bf Lower panel:} variation of the colors as a function of radius.}
\label{colorprofile}
\end{figure}

\begin{figure}
  \resizebox{0.7\hsize}{!}{\includegraphics[angle=-90]{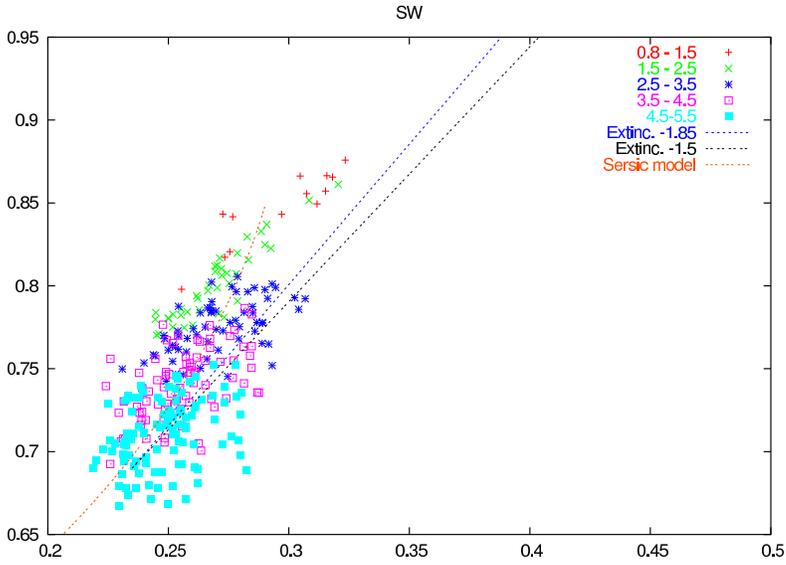}}
\caption{$J-H$ vs $H-Ks$ diagram for colors measured at positions between 0.8 
and 5.5 arcsec from the center of NGC~1097, constructed for the South-West 
quadrant. Diagrams for other quadrants are similar. Colors are binned into
five radial ranges, and coded  with different symbols. Each 
range (in arcsec), together with its corresponding symbol is shown in the 
top-right corner of the diagram. Two ``standard'' extinction curves (see 
Section 3.2) are plotted with black and blue lines. With the red line we
plot the colors for the Sersic $n=2$ model, which we used to fit the radial 
profile.} \label{sw}
\end{figure}

\begin{figure}
\begin{minipage}{0.7\hsize}
\includegraphics[width=0.7\textwidth]{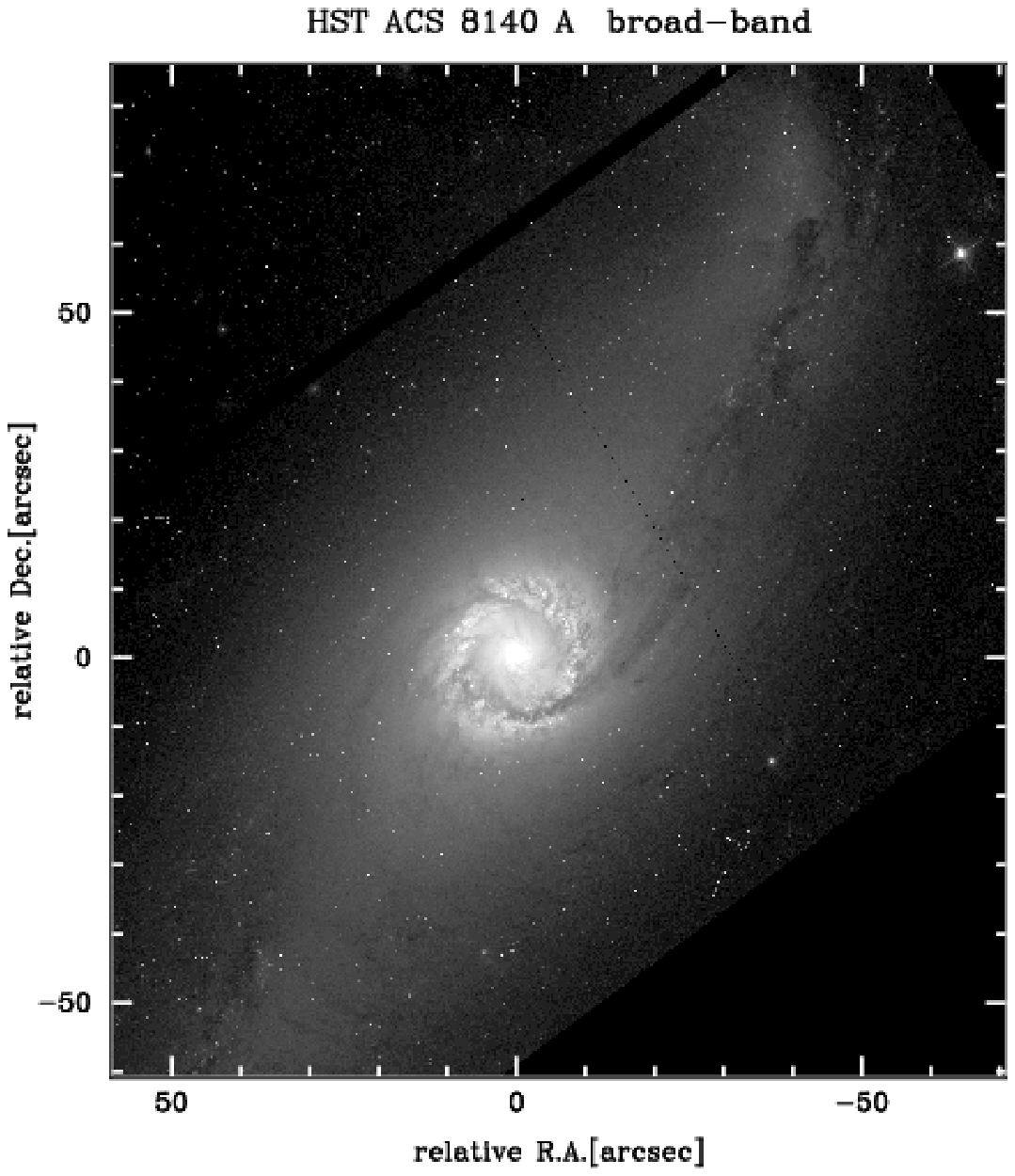}
\label{acs}
\end{minipage}

\label{color}
\includegraphics[width=0.5\textwidth]{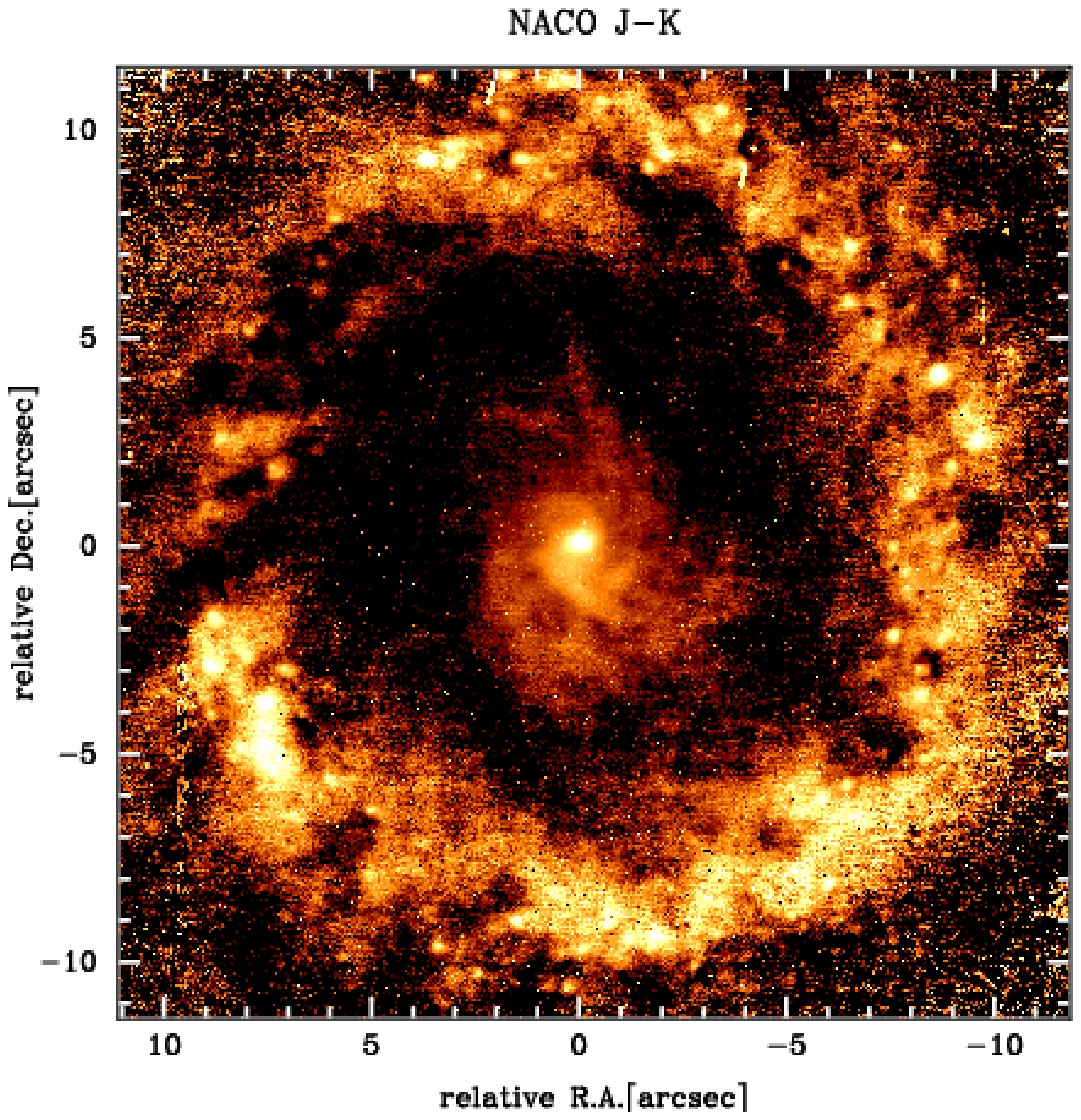}
\vspace{0.cm}\includegraphics[width=0.515\textwidth]{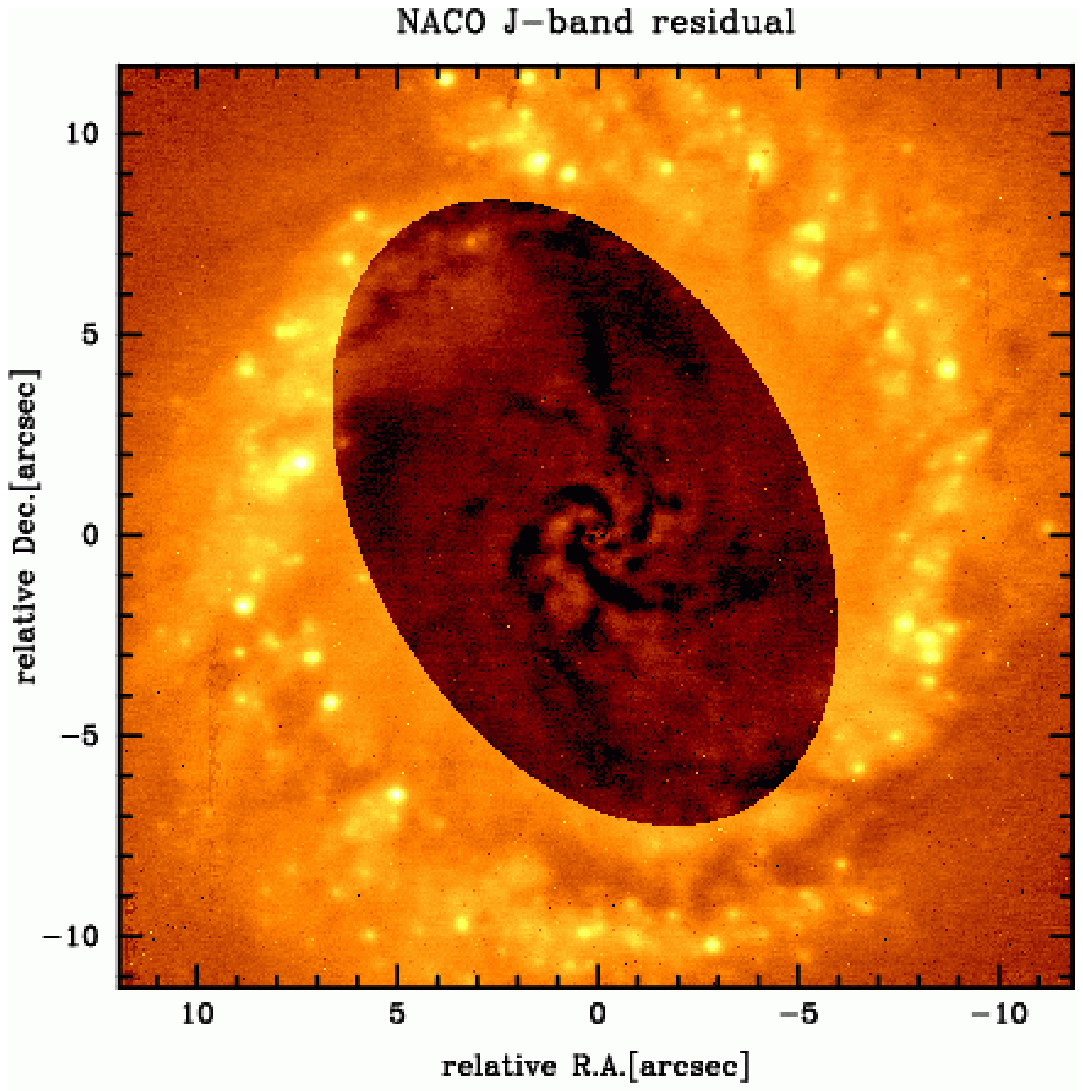}
\label{ellipse-residual-j-extract}

\vspace{0.cm}\caption{{\bf Top:} HST/ACS-F814W panoramic view of NGC~1097. The image is in
logarithmic scale, and greyscale cuts are chosen to emphasize the large-scale 
dust lanes and the nuclear star-forming ring. 
{\bf Bottom-left:} NACO $J-Ks$ color image
of the central $\sim 1.7 \times 1.7$ kpc region of NGC~1097, showing the 
nucleus, the central spiral arms extending up to 400 pc from the center, and 
the star-forming ring, saturated in this image. 
{\bf Bottom-right:} NACO $J-$band image of central region  of NGC~1097 after
subtraction of a simple ellipse model from the   oval region. 
The central spiral 
arms are now seen as dark channels, some extending up to the star-forming ring.
North is up, East is to the left in all panels.}

\end{figure}

\begin{figure}
\center
\vspace{-0.9cm}\includegraphics[width=0.6\textwidth]{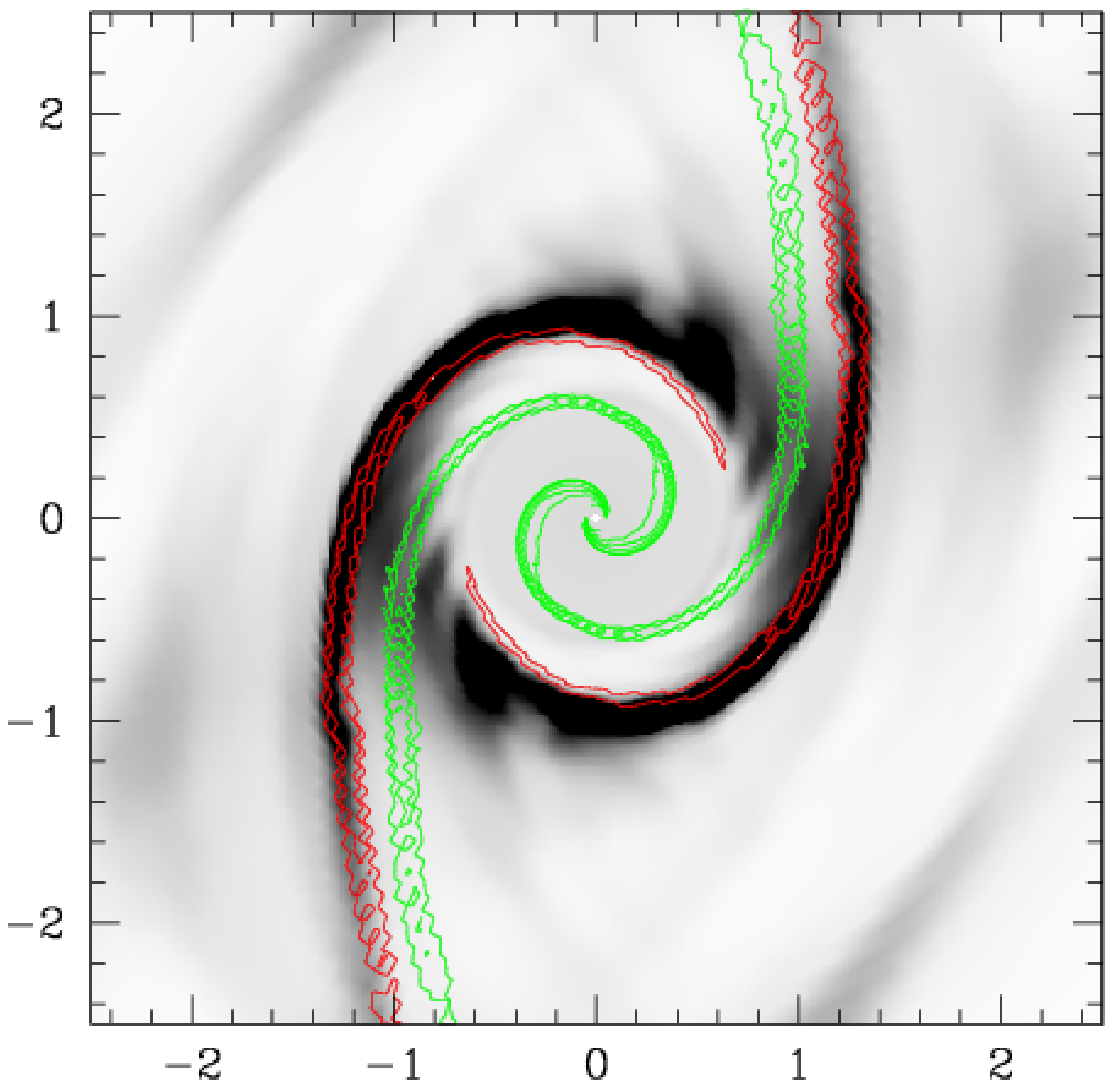}

\center
\hspace{-1.cm}\vspace{-.5cm}\includegraphics[width=0.7\textwidth]{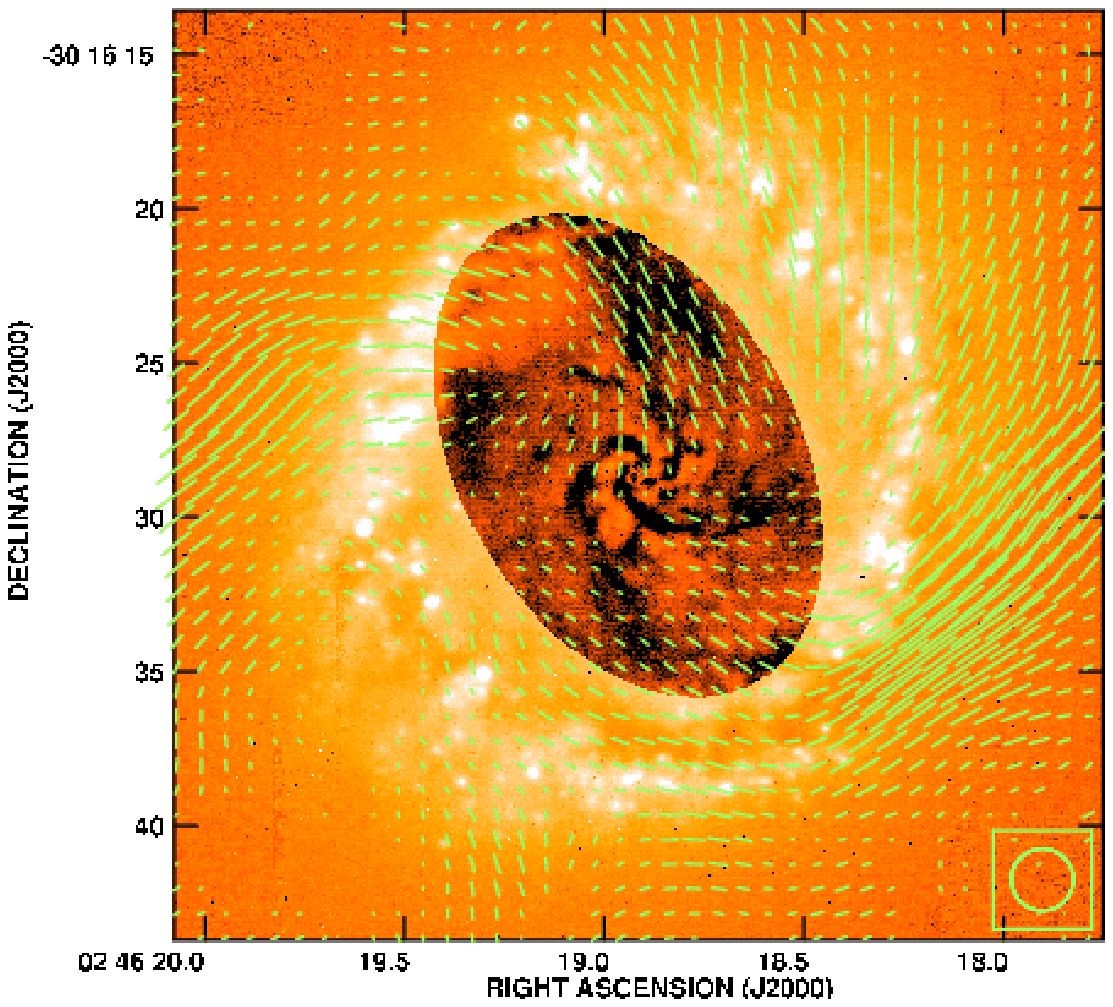}
\caption{\footnotesize{{\bf Top:} Illustration of the difference between 
dynamics of gas in the galactic plane and of the extraplanar gas, when the 
gas is subject to forces from a bar (vertical in the plot, 12-kpc long). 
Characteristics of gas flow are taken from models 8S05r 
and 8S20r by Maciejewski (2004b), with velocity dispersion in gas of 5 and 20 
\kms, respectively, at the evolutionary time of 180 Myr. For the model of low 
velocity dispersion in gas, typical for the in-plane gas, density is displayed 
in greyscale (darker shades represent higher density), and contours outlining 
shocks are drawn in red. Overplotted are contours in green, outlining shocks 
in the model with high velocity dispersion, representing the extraplanar gas. 
These models are built for a single bar, thus they do not reproduce 
the alignment of the central spiral with the inner bar. Units on axes are in 
kiloparsec.
{\bf Bottom:} Green lines, showing magnetic field vectors in the central 
$\sim 2 \times 2$ kpc of NGC~1097 (from polarized emission at 3.5 cm; Beck 
et al. 1999, 2005), are overplotted on the $J-$band 
residual image from the bottom-right panel of Fig. 5. 
The beam resolution of radio map  shown on bottom-right corner.}}

\end{figure}

\clearpage
\oddsidemargin=-1cm
\tabletypesize{\scriptsize}

\begin{table}
\begin{center}
  \begin{tabular}{llllll}
 \tableline\tableline
    Band & AGN & SB galaxy & n & r$_{eff}$ & r$_{eff}$\\
         & mag   & mag/arcsec   &   & \arcsec   & pc \\
 \tableline
    J & 12.1 &18 & 1.95 & 13.0& 890\\
    H & 10.8 & 16.2 & 2.05 & 11.5 & 800\\
    Ks& 9.8 &16.7  & 2.00 & 10.2 & 720\\
 \tableline
  \end{tabular}
  \caption{Parameters of the best fit to the surface brightness profile of 
the inner 3 arcsec ($\sim$210 pc) of NGC~1097. Col. 1: AGN magnitude; 
Col. 2: underlying galaxy  surface brightness within an aperture equal to the FWHM; 
Col. 3: Sersic exponent; Cols 4,5: effective radius.
    \label{fitparas}}
\end{center}
\end{table}

\clearpage

\begin{table}
\begin{center}
  \begin{tabular}{llllllll}
\tableline\tableline
    r      &  J  &H & Ks & (J-H)$_s$ & (H-Ks)$_s$ & (J-H)$_m$ & (H-Ks)$_m$\\
    \arcsec & \% & \% &\%   & mag       & mag       \\
\tableline
    0.1 & 67.7 & 75.9 & 90.9 & 0.88 & -0.14 & 0.95 & 0.29 \\
    0.2 & 45.2 & 51.8 & 68.4 & 0.94 & 0.27  & 0.93 & 0.29 \\
    0.3 & 36.3 & 41.4 & 59.3 & 0.92 & 0.23  & 0.92 & 0.29 \\
    0.4 & 28.0 & 32.0 & 48.9 & 0.91 & 0.25  & 0.91 & 0.29 \\
    0.5 & 23.0 & 26.4 & 42.3 & 0.90 & 0.26  & 0.89 & 0.28 \\
    0.6 & 18.8 & 21.7 & 36.1 & 0.80 & 0.27  & 0.89 & 0.28 \\
    0.7 & 15.8 & 18.4 & 31.6 & 0.88 & 0.27  & 0.88 & 0.28 \\
    0.8 & 13.5 & 15.7 & 27.7 & 0.87 & 0.27  & 0.87 & 0.28 \\
    0.9 & 11.6 & 13.6 & 24.3 & 0.86 & 0.28  & 0.86 & 0.28 \\
    1.0 & 10.1 & 11.9 & 21.6 & 0.86 & 0.28  & 0.86 & 0.28 \\
    2.0 &  3.9 &  4.7 &  9.4 & 0.82 & 0.28  & 0.81 & 0.27 \\
\tableline
    \end{tabular}
  \caption{Point source contribution and integrated colors of the
    underlying galaxy 
at radii between 0.1 and 2 arcsec. Colors with subindex $_s$ are measured after
subtraction of the nucleus  emission, while those with $_m$ are derived from the 
Sersic model directly.
    \label{fitres} }
\end{center}
\end{table}

\end{document}